\documentclass[a4paper,fleqn,usenatbib]{mnras}

\usepackage{mathptmx}

\usepackage[T1]{fontenc}
\usepackage{ae,aecompl}

\usepackage{graphicx}   
\usepackage{amsmath}    
\usepackage{amssymb}    
\usepackage{natbib}

\title[Three-lane and multi-lane signatures]{Three-lane and multi-lane signatures of planets in planetesimal
disks}

\author[T. V. Demidova \& I. I. Shevchenko]{
Tatiana V. Demidova,$^{1}$\thanks{E-mail: proxima1@list.ru}
Ivan I. Shevchenko$^{1,2}$
\\
$^{1}$Pulkovo Observatory of the Russian Academy of Sciences, Pulkovskoje Avenue 65, St. Petersburg 196140, Russia\\
$^{2}$Lebedev Physical Institute of the Russian Academy of Sciences, 53 Leninskiy Prospekt, Moscow 119991 , Russia
}

\date{Accepted XXX. Received YYY; in original form ZZZ}

\pubyear{2016}

\begin{document}
\label{firstpage}
\pagerange{\pageref{firstpage}--\pageref{lastpage}}
\maketitle

\begin{abstract}
In massive numerical experiments we show that a planet embedded in
a planetesimal disk induces a characteristic multi-lane
``planetosignature'' representing a pattern of several
stellar-centric rings. If the planet's mass is large enough, the
multi-lane signature degenerates to a three-lane one: then it
consists of three rings, one bright coorbital with the planet, and
two dark gaps in the radial distribution of the particles. The
gaps correspond to orbital resonances 2:1 and 1:2 with the planet.
This theoretical prediction may explain recent ALMA
observations of the disk of HL~Tau.
\end{abstract}




\section{Introduction}

The motion of massive bodies inside a planetesimal disk may
produce structures stable on secular time scales. In
a previous study \cite{DS15}, we explored how a secular
single-arm spiral pattern emerged in a circumbinary planetesimal
disk devoid of planets. Here we explore models with a planet
embedded in a circumbinary disk; namely, we investigate the
formation of patterns in such a disk, and compare the emerging
patterns with those arising in the case when the central star is
single.

\section{Model}

We model disks for two cases of the mass ratio of the central
stellar binary: a circular binary with masses $m_1=M_{\odot}$ and $m_2=0.2 M_{\odot}$ (model~1) and a circular binary with masses $m_1=m_2=M_{\odot}$ (model~2). For comparison, we take the model of a single central star with mass $m=1.2M_{\odot}$ (model~3). The orbital periods of the binaries in models 1 and 2 are both set to $P_b=0.2$~yr. The planet of Jovian mass is put initially in a circular orbit around the stellar barycenter. Its orbital radius corresponds to mean motion resonances with the binary, the ratios of the orbital periods of the planet and the binary set equal to $5:1$, $11:2$, $6:1$, $13:2$, $7:1$, $15:2$, $8:1$. The choice of this grid of locations is motivated by the existing data that the {\it Kepler} circumbinary planets are mostly located in resonance cells at the outer border of the chaotic region around the central binary (see~\citealt{PS13,PS16}; in particular, Table~3
in~\citealt{PS16}). In model~3 (the host star is single), the
planet's orbit is started at the same grid of radial distances as in model~1. The disk consists of 20000 massless (passively
gravitating) planetesimals  initially distributed from $0.3$ to
$5.3$~AU in the radial distance from the barycenter in such a way that the surface density $\Sigma$ decreases with the radial
distance $r$ according to the law $\Sigma \propto r^{-1}$. In each model, the motion was computed for $5\cdot10^4$~yr.

\section{Patterns}
\begin{figure}
\begin{center}
\includegraphics[width=1\columnwidth]{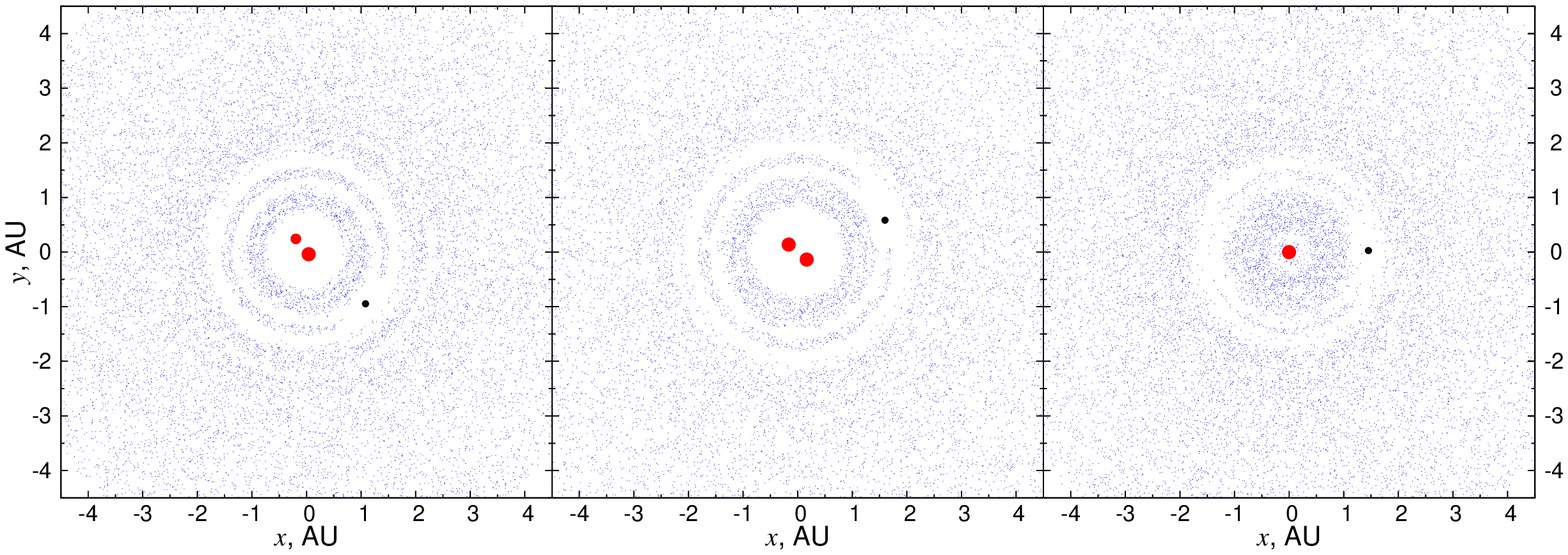}
\caption{\normalsize The evolved distributions  of planetesimals
in models~1, 2, and 3 (from left to right).}
\label{f1}
\end{center}
\end{figure}

The performed simulations make evident an emerging ring-shaped
pattern (consisting of the planetesimals in the ``tadpole'' and
``horse-shoe'' orbits), coorbital with the planet (see
Fig.~\ref{f1}, where the case of planetary resonance 8:1 with the central binary is illustrated). They are most pronounced in
models~1 and 2. For a single host star (as in model~3), similar
coorbital patterns were revealed in~\cite{Oetal00,QT02,KH03}.

To estimate the pattern appearance quantitatively, we calculate
the surface density distribution along the radius from the
barycenter: the planetesimal disk is subdivided into 320 annular
bands with the radial step $0.02$~AU, then the number of particles in each band is calculated and divided by the band area (thus, the local surface density is calculated as a function of the radial distance).

As revealed by \cite{W80}, a chaotic band around the orbit
of a small-mass body (planet) moving around a main body (star)
exists, due to the overlap of first-order mean motion resonances
$(p+1):p$ with the planet on the inner side of the planet's orbit, and resonances $p:(p+1)$ on the outer side of the orbit (integer numbers $p \gg 1$). If $\mu \equiv m_2/(m_1+m_2) \ll 1$ (where $m_1$ and $m_2$ are the masses of the planet and the star, respectively), in the planar circular restricted three-body problem the radial half-width of this band is given by

\begin{equation} \Delta a_\mathrm{Wisdom} \approx 1.3 \mu^{2/7}
a_\mathrm{p} \label{Wisdom} \end{equation}

\noindent \citep{Detal89}, where $a_\mathrm{p}$ is the semimajor
axis of the planet's orbit, or

\begin{equation} \Delta a_\mathrm{Wisdom} \approx 1.57 \mu^{2/7}
a_\mathrm{p} \label{Wisdom_MD} \end{equation}

\noindent \citep{MD99}. Further on, we use the latter formula,
because its validity was verified in \cite{MD99} in massive model simulations.

On the other hand, the radial half-width of the coorbital (with
the planet) band of stable horse-shoe and tadpole orbits can be
estimated as approximately equal to the radius of the Hill sphere (for an illustration see figure~3.28 in \citealt{MD99}). In the planar circular restricted three-body problem the Hill radius is given by

\begin{equation}
\Delta a_\mathrm{Hill} \approx \left( \frac{\mu}{3} \right)^{1/3}
a_\mathrm{p} \approx 0.693 \mu^{1/3} a_\mathrm{p} \label{Hill}
\end{equation}

\noindent (see, e.g., \citealt{MD99}).

From equations (\ref{Wisdom_MD}) and (\ref{Hill}) one has

\begin{equation}
\frac{\Delta a_\mathrm{Wisdom}}{\Delta a_\mathrm{Hill}} \approx
2.26 \mu^{-1/21} . \label{WH}
\end{equation}

\noindent Therefore, the ratio $\Delta a_\mathrm{Wisdom} / \Delta a_\mathrm{Hill}$ is very insensitive to the mass parameter: in the range of $\mu$ from 0.0005 (model~2) to 0.01 it is $\approx 3$, changing only slightly, from 3.25 to 2.81. (In model~1, where $\mu \approx 0.0008$, one has $\Delta a_\mathrm{Wisdom} / \Delta a_\mathrm{Hill} \approx 3.17$.) Therefore, in this broad range of $\mu$, the radial extent of the material kept coorbital with the planet is expected to be about one third of the radial extent of the theoretical Wisdom gap. In other words, the central one third (in radial extent) of the Wisdom gap may contain stable horse-shoe and tadpole material.

Therefore, from the theoretical viewpoint, one may expect
existence of a ring-like pattern, surrounding the orbit of the
planet embedded in the planetesimal disk. This pattern consists of at least three lanes: the bright central one (which we designate further on as Bc, i.e., bright central, or bright coorbital), and two components of the broader Wisdom gap (which we designate further on as Dc$^\mathrm{int}$ and Dc$^\mathrm{ext}$, i.e., dark central, or dark coorbital, internal and external). These two components arise due to the division of the theoretical Wisdom gap by the Bc lane.

\begin{figure}
\begin{center}
\includegraphics[width=1\columnwidth]{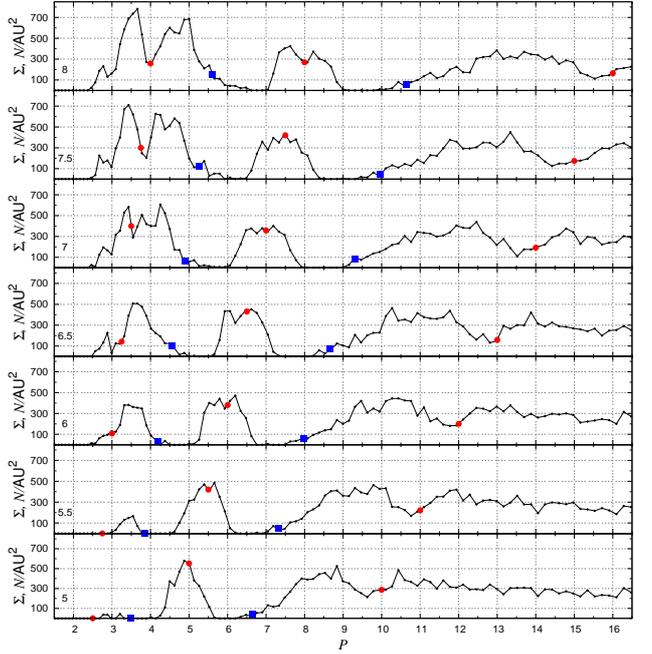}
\caption{\normalsize The local surface density as a function of
the planet's orbital period (in units of the host binary's period) in model~1. Resonance of the planet with the host binary is specified at the left bottom corner of each panel. The red dots indicate the location of mean motion resonances 2:1, 1:1, 1:2 of planetesimals with the planet. The blue squares indicate the location of the Wisdom gap borders, according to
formula~(\protect\ref{Wisdom_MD}).}
\label{f2}
\end{center}
\end{figure}

\begin{figure}
\begin{center}
\includegraphics[width=1\columnwidth]{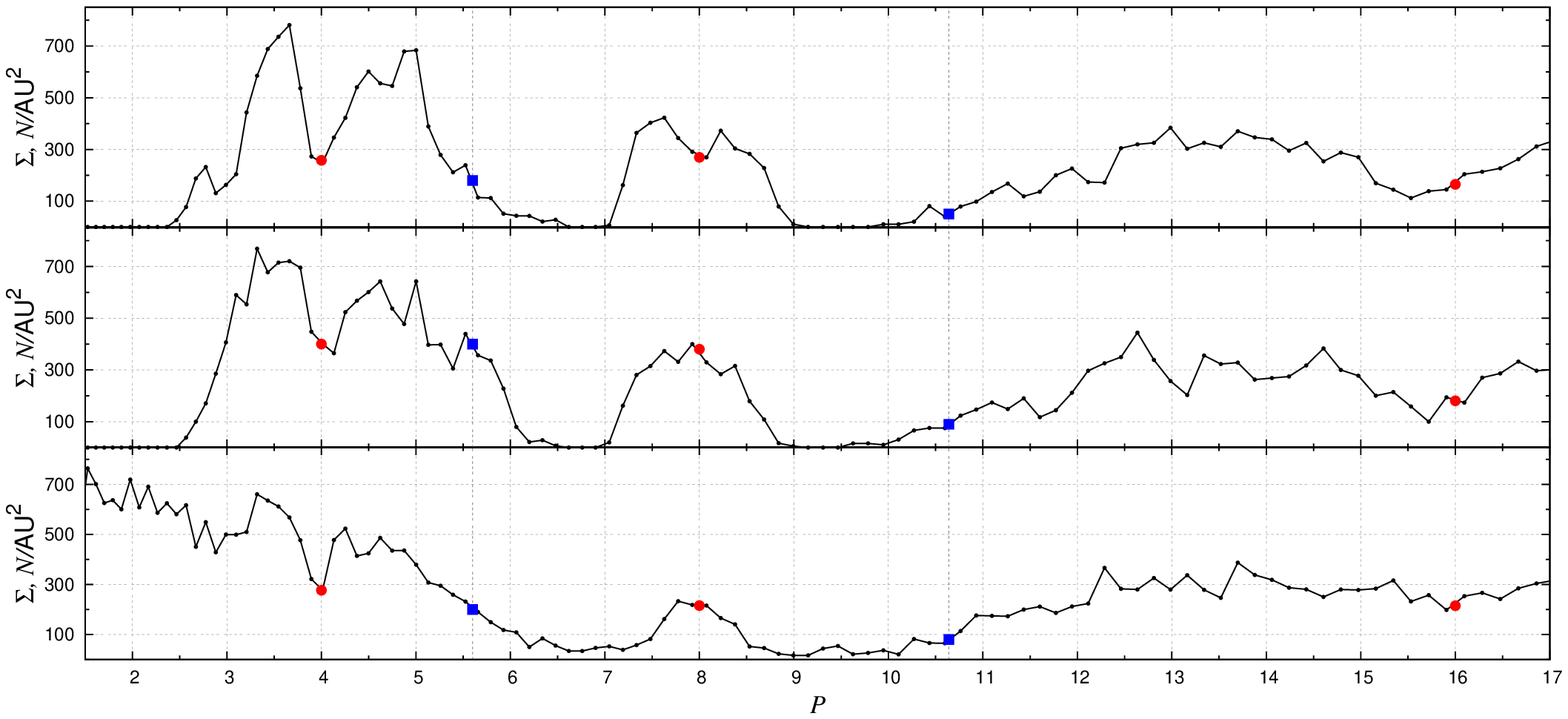}
\caption{\normalsize The local surface density as a function of
the planet's orbital period (in units of the host binary's period) in models~1, 2 and 3 (from top to bottom). The first two panels correspond to the case of resonance 8:1 of the planet with the host binary. The red dots indicate the location of mean motion resonances 2:1, 1:1, 1:2 of planetesimals with the planet. The blue squares indicate the location of the Wisdom gap borders, according to formula~(\protect\ref{Wisdom_MD}).}
\label{f3}
\end{center}
\end{figure}

As follows from Fig.~\ref{f2}, where the results of
simulations in model~1 are presented, this three-lane complex is
present in all panels. In each panel, the disk density profile has a strong peak at resonance 1:1 with the planet (this peak
corresponds to the Bc lane), as well as prominent gaps
corresponding to the Dc$^\mathrm{int}$ and Dc$^\mathrm{ext}$
lanes. This {\it planetosignature} is observed for all choices of the initial radial location of the planet.

What is more, in each panel one observes additional pronounced
minima, which are located at mean motion resonances 2:1 and 1:2
with the planet. We designate the corresponding lanes as D$_{2:1}$ and D$_{1:2}$ (the dark lanes at resonances 2:1 and 1:2). They are separated from the central three-lane complex
Dc$^\mathrm{int}$--Bc--Dc$^\mathrm{ext}$ by two (internal and
external) bright lanes, which we designate as Bb$^\mathrm{int}$
and Bb$^\mathrm{ext}$ (``bright barriers'').

In total, one has a multi-lane (in fact, seven-lane) complex:
D$_{2:1}$--Bb$^\mathrm{int}$--Dc$^\mathrm{int}$--Bc--Dc$^\mathrm{ext}$--Bb$^\mathrm{ext}$--D$_{1:2}$.

In models~2 and 3 (Fig.~\ref{f3}), the central three-lane complex
appears to be basically the same. However, the D$_{2:1}$ and
D$_{1:2}$ lanes are prominent only in the models with large enough
planetary orbits. Therefore, depending on the system parameters,
the observed pattern can be either three-lane or seven-lane. The
three-lane pattern can arise, instead of the generic seven-lane
pattern, in two cases: (1)~simply because the 2:1 and 1:2
resonances are not prominent, (2)~if the D$_{2:1}$ and D$_{1:2}$
lanes overlap, respectively, with the Dc$^\mathrm{int}$ and
Dc$^\mathrm{ext}$ lanes (thus, the ``bright barriers'' Bb vanish).
In the second case, the basic minima in the particle distribution
shift to resonances 2:1 and 1:2 (whereas the Bc lane broadens);
therefore, the observable pattern can be designated as
D$_{2:1}$--Bc--D$_{1:2}$.

Let us derive an approximate condition for the second case. This
can be done by equating the Wisdom gap borders' locations (given
by equation~(\ref{Wisdom_MD})) to the locations of resonances 2:1
and 1:2 (given by Kepler's third law). One has two equations, for
the inner and outer borders: $1 - 1.57 \mu^{2/7} = 2^{-2/3}
\approx 0.63$ and $1 + 1.57 \mu^{2/7} = 2^{2/3} \approx 1.59$.
Solving them, one has two approximate critical values of $\mu$:
0.006 and 0.03, respectively. They are very approximate, as
Wisdom's theory is suitable at small values of $\mu$ only (see
figure~2 in \citealt{S15}). Therefore, we adopt the critical $\mu$
value $\sim 0.01$. At such values of $\mu$ one expects the
degeneration of the seven-lane complex into the three-lane one.

For a single host star, rather pronounced gaps at resonance 2:1
and at resonance 1:2 (in separate models) with the planet of 5
Jovian masses were found recently in~\cite{TW16}. The gaps
originated on the time scale of $10^5$~yr. Here we find that the
opening of the resonant gaps D$_{2:1}$ and D$_{1:2}$ is much more
pronounced in circumbinary disks than in a disk of a single host
star.

The reason seems to be as follows: the presence of a stellar
companion induces more secondary resonances in the neighborhood of
both resonances 2:1 and 1:2, and their overlap causes more chaos
in the phase space of motion. However note that, on the other
hand, the stellar companion presence also induces precession of
the planetary and planetesimal orbits, and in case of fast
differential precession the splitting of resonances might be great
enough to prevent the overlap; for an analysis of similar
situations see \cite{Meal14}.

The emergence of resonant gaps D$_{2:1}$ and D$_{1:2}$ themselves
is theoretically expectable, as both resonances 2:1 and 1:2 may
induce large zones of instability in the phase space of motion,
depending on initial conditions and system parameters
(see~\citealt{M02}).

The circumbinary central cavities evident in the density profiles
and in the snapshots of the evolved circumbinary disks are a
natural outcome of the disk evolution, because a large central
chaotic circumbinary zone exists at all eccentricities of the
planetesimal if $\mu \gtrsim 0.05$ \citep{S15}, where $\mu = m_2
/(m_1 + m_2)$ is the mass parameter of the binary. According to
empirical criteria derived by \cite{HW99}, the expected radii of
the central chaotic zone should be $\approx 2.2 a_\mathrm{b}$ and
$\approx 2.4 a_\mathrm{b}$ in models 1 and 2, respectively. This
agrees well with the results of our simulations: the innermost
maxima of the particle distributions in Figs.~\ref{f2} and
\ref{f3} are situated basically at $a/a_\mathrm{b} \approx 2.3$
($a/a_\mathrm{b} = (P/P_\mathrm{b})^{2/3}$, where $P/P_\mathrm{b}
\approx 3.5$).

In the density profiles, the 1:1 peak corresponds to the ring-like
pattern coorbital with the planet (i.e., the Bc lane). To
characterize the ``survivability'' of the Bc lane, we compute the
surface density $\Sigma$ (averaged over the ring) as a function of
time, in all three models. The radial boundaries of the Bc lane in
models 1 and 2 are set at the radii where the local surface
density $\Sigma$ goes to zero, and the boundaries in model~3 are
chosen to be the same as in model~1. The surface density $\Sigma$
is defined as the number of particles moving inside the ring
boundaries, divided by the ring area.

The problem on the stability of planetesimal motion in
protoplanetary disks of multiple stars was discussed earlier
in~\cite{VE06,VE07,VE08}. Here we find that the survivability of
the coorbital ring pattern turns out to be rather different in
disks of binary and single host stars: unexpectedly enough, it
seems to be much greater in the binary case. This is illustrated
in Table~\ref{tab:surv}: in model~3, the percentage of surviving
particles is $\sim 2$ times less than in the first two models.

\begin{table}
\begin{center}
\caption{The ratio of final and initial populations of the
coorbital ring}
\begin{tabular}{| c | c | c | c | c | c | c | c |}
\hline
$P_p/P_b$  &  $5$  &  $5.5$  &  $6$ & $6.5$ & $7$ & $7.5$
& $8$   \\
\hline
Model 1 & $0.907$ & $0.916$ & $0.949$ & $0.947$ & $0.906$ & $0.942$ & $0.942$  \\
\hline
Model 2 & $0.916$ & $0.932$ & $0.928$ & $0.928$ & $0.931$ & $0.908$ & $0.920$  \\
\hline
Model 3 & $0.585$ & $0.521$ & $0.494$ & $0.468$ & $0.388$ & $0.433$ & $0.448$ \\
\hline
\end{tabular}
\end{center}
\label{tab:surv}
{\small Note: The initial and final populations
correspond to time $t=1000$~yr and $t=50000$~yr, respectively.}
\end{table}

\section{Application to the HL~Tau disk}

Planet-like ``clumps'' and coorbital and neighbouring ring-like
patterns are directly observed now in circumstellar disks
\citep{Cetal16}. Our inference that a multi-lane pattern can be
generated by a planet embedded in a planetesimal disk may
explain recent ALMA observations of the disk of HL~Tau.

As follows from figure~3 in~\cite{Cetal16}, dark ring-like
features D1 and D2 are situated at relative radii $\approx 0.63$
and $\approx 1.60$ (if the radius of the main bright feature B1,
that with a planet-like ``clump'', is set to unity). These numbers
practically coincide with $2^{-2/3} \approx 0.63$ and $2^{2/3}
\approx 1.59$, corresponding to mean motion resonances 2:1 and 1:2
with the clump. Therefore, they correspond to the D$_{2:1}$ and
D$_{1:2}$ lanes in our models. Why the pattern here is three-lane,
not the generic seven-lane?

Arguing that the emission from the clump is dominated by thermal
dust emission, \cite{Cetal16} estimate a dust mass in the clump in
the range 3--8 Earth masses. Setting the dust-to-gas ratio equal
to the standard value 1/100, one obtains 1--3 Jovian masses for
the full mass of the clump. On the other hand, the mass of the
HL~Tau star is $\approx 0.55 M_{\odot}$ \citep{Beal90}. Therefore,
the mass parameter of the star-clump system can be estimated as
$\mu \sim 0.002$--0.006. As discussed in the previous Section, the
D$_{2:1}$--Bc--D$_{1:2}$ three-lane pattern is expected to emerge
at $\mu \gtrsim 0.01$. Taking into account that the both estimates
are approximate, we conclude that the D$_{2:1}$--Bc--D$_{1:2}$
interpretation of the observed pattern is rather plausible.

Let us note that other interpretations of the HL~Tau disk
structure may exist. \cite{Teal15} introduced multiple-planet
models with two inner planets located in the D1 and D2 gaps in the
HL~Tau disk, thus explaining the opening of the gaps by the
purging effect of the planets.

When applying our model inferences to HL~Tau, it is important
to note that the single-star case corresponds to our model~3. In
the bottom panel of Fig.~\ref{f3} we see that the 2:1 and 1:2
patterns are less pronounced in model~3 than in models 1 and 2
(corresponding to the circumbinary case and represented in the
upper panels). On the other hand, the mass parameter adopted in
all our models is rather low ($\mu \sim 0.001$); and, as noted
above, according to simulations by \cite{TW16} for a planet of 5
Jovian masses (i.e., 5 times greater than in our model~3) orbiting
around a single star, the 2:1 and 1:2 gaps generated in the disk
are rather pronounced.

It is also important to point out that our model disks are
planetesimal, whereas the HL~Tau disk is gas-dust. The presence of
gas may affect the process of forming the disk patterns. The dust
experiences aerodynamic drag, and the disk's viscosity refills the
opening gaps \citep{GT80}. The formation of gaps can be retarded
or they may be smoothed. In future it would be important to
consider how the dissipation effects may alter the presented
picture.

In this connection we note that there exist various estimates of
the HL~Tau system age: from $\sim 0.2$~Myr~\citep{Geal11} to $\sim
1$~Myr~\citep{Beal90}; and this is 4--20 greater than our typical
simulation time (0.05~Myr). However, it should be taken into
account that the mass of HL~Tau is 2.2 times less than the stellar
mass in our model~3. This means that the timescales measured in
planetary periods are 1.5 times less different: they are in the
range 2.7--13. According to Table~1, the Bc lane in the single
star model is depleted rather quickly; therefore, our model~3
result may favour low values for the HL~Tau system age. To judge
whether this is the case, future integrations on longer timescales
are required, as well as taking the gas presence into account.

\section{Conclusions}

We conclude that any observed presence of ring-like patterns in
circumbinary disks may betray the existence of a planet that
``shepherds'' the ring. What is more, the planet's mass can
be estimated by measuring the radial extents of the pattern
components.

Our model simulations show that a planet embedded in a
planetesimal disk may induce a characteristic multi-lane
``planetosignature'' --- a pattern consisting of several
stellar-centric rings. The planetary multi-lane pattern
seems to reveal itself most definitely in circumbinary disks.

If the planet's mass is large enough ($\mu \gtrsim 0.01$), the
multi-lane signature degenerates to a three-lane one: then it
consists of three rings, one bright coorbital with the planet, and
two dark gaps in the radial distribution of the particles. The
gaps correspond to orbital resonances 2:1 and 1:2 with the planet.
This theoretical prediction may explain recent ALMA
observations of the disk of HL~Tau.

Finally, we note that the currently used general definition of a
planet (in the definition's part requiring the planet to purge a
neighborhood of its orbit \citealt{IAU06}) may need a
reformulation, if one wishes the circumbinary planets to be
encompassed by this definition, because the survivability of the
coorbital planetesimal ring seems to be outstanding for such
planets.

\section*{Acknowledgements}

It is a pleasure to thank the referee for most valuable and useful
remarks. This work was supported in part by the Russian Foundation
for Basic Research (projects Nos.\ 14-02-00319 and 14-02-00464)
and the Programmes of Fundamental Research of the Russian Academy
of Sciences ``Fundamental Problems of Nonlinear Dynamics'' and
``Fundamental Problems of the Solar System Study and
Exploration''. The computations were partially carried out at the
St.~Petersburg Branch of the Joint Supercomputer Centre of the
Russian Academy of Sciences.



\bibliographystyle{mnras}
\bibliography{biblio}

\label{lastpage}
\end{document}